\documentstyle[12pt]{article}
\let\LARGE=\Large
\let\Large=\large
\let\large=\normalsize
\newcommand{\be}[3]{\begin{equation}  \label{#1#2#3}}     
\newcommand{\ee}{ \end{equation}}
\newcommand{\ba}{\begin{array}}
\newcommand{\ea}{\end{array}}
\newcommand{\NP}[3]{{\em Nucl. Phys.}{ \bf B#1#2#3}}
\newcommand{\PRD}[2]{{\em Phys. Rev.}{ \bf D#1#2}}
\newcommand{\PRL}[2]{{\em Phys. Rev. Lett.}{ \bf #1#2}}

\newcommand{\ft}[2]{{\textstyle\frac{#1}{#2}}}

\def\beq{\begin{equation}}
\def\eeq{\end{equation}}
\def\beqa{\begin{eqnarray}}
\def\eeqa{\end{eqnarray}}

\renewcommand{\d}{\delta}
\newcommand{\pa}{\partial}

\newcommand{\m}{\mu}

\newcommand{\n}{\nu}

\begin{document}
\thispagestyle{empty}
\rightline{HUB-EP-98/4}
\rightline{THU-98/02}
\rightline{QMW-PH-98-01}\vspace{1mm}
\rightline{hep-th/9801081}
\vspace{2truecm}
\centerline{\bf \LARGE
Higher-Order Black-Hole Solutions}

\centerline{\bf \LARGE in $N=2$ Supergravity and Calabi-Yau
String Backgrounds} 
\vspace{1.2truecm}
\centerline{\bf 
Klaus Behrndt$^1$\footnote{behrndt@qft2.physik.hu-berlin.de}\ , \ 
Gabriel Lopes Cardoso$^2$\footnote{cardoso@fys.ruu.nl}\ , \
Bernard de Wit$^2$\footnote{bdewit@fys.ruu.nl}\ , }

\centerline{\bf
Dieter L\"ust$^1$\footnote{luest@qft1.physik.hu-berlin.de}
\ , \ Thomas Mohaupt$^3$\footnote{mohaupt@hera1.physik.uni-halle.de}
 \  and  \  Wafic A. Sabra$^4$\footnote{sabra@qft2.physik.hu-berlin.de}}
\vspace{.5truecm}
{\em 
\centerline{$^1$Humboldt-Universit\"at, Institut f\"ur Physik,
D-10115 Berlin, Germany}

\centerline{$^2$Institute for Theoretical Physics, Utrecht University,
3508 TA Utrecht, Netherlands}

\centerline{$^3$Martin-Luther-Universit\"at Halle-Wittenberg, 
Fachbereich Physik,
D-06122 Halle, Germany}

\centerline{$^4$Physics Department, QMW, Mile End Road, London E1 4NS,
United Kingdom}}

\vspace{1.0truecm}
\begin{abstract}
Based on special geometry, we consider corrections to
$N=2$ extremal black-hole solutions and their 
entropies originating from higher-order derivative terms in 
$N=2$ supergravity. These corrections are described by a holomorphic
function, and the higher-order
black-hole solutions can be expressed in terms of symplectic
Sp(2$n$+2) vectors. We apply the formalism to
$N=2$ type-IIA Calabi-Yau string compactifications and compare 
our results to recent related results in the literature.
\end{abstract}
\vspace{1cm}

{\centerline {\it Dedicated to the memory of Constance Caris}}

\bigskip \bigskip
\newpage


One of the celebrated successes within the recent non-perturbative
understanding of string theory and M-theory is the matching of the
thermodynamic Bekenstein-Hawking black-hole entropy with the
microscopic entropy based on the counting of the relevant D-brane
configurations which carry the same charges as the black hole
\cite{StromVafa, CallanMal}.  This comparison works nicely for
type-II string, respectively, M-theory backgrounds which break half of the
supersymmetries, i.e.  exhibit $N=4$ supersymmetry in four
dimensions. In this paper we will discuss charged black-hole solutions
and their entropies in the context of $N=2$ supergravity in four
dimensions.  Four-dimensional $N=2$ supersymmetric vacua are obtained
by compactifying the type-II string on a Calabi-Yau threefold $CY_3$ 
or by M-theory compactification on $CY_3\times S^1$. In addition, the
heterotic string on $K3\times T^2$ also leads to $N=2$ supersymmetry
in four dimensions and the heterotic and the type-II vacua are 
expected to be related by string-string duality \cite{KachruVafa}.

Extremal, charged, $N=2$ black holes, their entropies and also the
corresponding brane configurations were discussed in several recent
papers 
\cite{FerrKallStro,BCDWKLM,rey,BehrndtLuestSabra, BehLu,msw,vafa,sen}.  
The macroscopic
Bekenstein-Hawking entropy ${\cal S}_{\rm BH}$, 
i.e. the area $A$ of the black-hole horizon,
can be obtained as the minimum of the graviphoton charge
$Z$ 
\cite{FerrKallStro}, 
\beqa
{\cal S}_{\rm BH}= {A\over 4}=\pi |Z_{\rm min}|^2\, , \label{area}
\eeqa
where, as a solution of the minimization procedure, the entropy as well as
the scalar fields (moduli) at the horizon depend only on the 
electric and magnetic black hole charges, but not on 
the asymptotic boundary values of the moduli fields. 
This procedure has been used extensively
to determine the entropy of $N=2$ black holes for type-II compactifications
on Calabi-Yau threefolds \cite{BCDWKLM}. One of 
the key features of extremal $N=2$ black-hole solutions
is that the moduli depend in general on $r$, but show a fixed-point
behaviour at the horizon. This fixed-point behaviour is implied 
by the fact that, at the  
horizon, full $N=2$ supersymmetry is restored; at the 
horizon the metric is equal to the Bertotti-Robinson 
metric, corresponding to the $AdS_2\times S^2$ geometry. This metric
can be described by a $(0,4)$ superconformal field theory 
\cite{Strominger}. The extremal black hole can be 
regarded as a soliton solution which interpolates between
two fully $N=2$ supersymmetric vacua, namely corresponding to $AdS_2\times 
S^2$ at the horizon and flat Minkowski spacetime at spatial 
infinity. 

The $N=2$ black holes together with their entropies which were 
considered so far, appeared as solutions of the equations of 
motion of 
$N=2$ Maxwell-Einstein supergravity action, where the
bosonic part of the action contains terms with
at most two space-time derivates (i.e., the Einstein action, 
gauge kinetic terms and the scalar non-linear $\sigma$-model). 
This part of the $N=2$ 
supergravity action can be encoded in a holomorphic 
prepotential $F(X)$, which is a function of the scalar fields $X$ 
belonging to the vector multiplets. However, the $N=2$ effective action 
of strings and M-theory contains in addition an infinite number 
of higher-derivative terms involving higher-order products of 
the Riemann tensor and the vector field strengths.
A particularly interesting subset of these couplings in $N=2$ 
supergravity can be again 
described by a holomophic function $F(X,W^2)$ 
\cite{BergshoeffdeRoodeWit,AntonGavaNarTayl,Bershadskyetal,deWit,DWCLMR,Moore}
, 
where the additional chiral superfield $W$ is the Weyl 
superfield, comprising 
the covariant quantities of conformal supergravity.
Its lowest component is the graviphoton field 
strength (in the form of an auxiliary tensor field $T_{\mu \nu}^-$), while 
the Weyl 
tensor appears at the $\theta^2$ level.
The aim of this paper is, using the superconformal calculus, to study the
$N=2$ black-hole solutions and the corresponding entropies for higher
order $N=2$ supergravity based on the holomorphic function $F(X,W^2)$.
We treat $W^2$ as a new chiral background and expand 
the black-hole solutions as a power series in $W^2$.
As an interesting example we compute the entropy of the charged $N=2$
black holes in type-II compactifications on a Calabi-Yau threefold.
At the end we compare our results to a recent computation of the 
microscopic Calabi-Yau black hole in M-theory \cite{msw,vafa}.

Let us start by recalling the $N=2$ black-hole solutions and 
their entropies in the case where the holomorphic function $F(X)$ 
does not depend on the Weyl multiplet. 
The bosonic $N=2$ supergravity action coupled to $n$ vector multiplets is 
given by
\beqa
S_{N=2}= \int {\rm d}^4x\, \sqrt{-g}\Big[-\ft12 R + g_{A \bar B}\,
\partial^{\mu} z^A \partial_{\mu} \bar z^{B} -\ft18i \left(\bar
{\cal N}_{I J} {F}^{- I}_{\mu \nu} {F}^{- J {\mu \nu}} \, - \, {\cal
N}_{IJ} {F}^{+ I}_{\mu \nu} { F}^{+ J {\mu \nu}}\right)\Big] ,
\label{action} 
\eeqa
where the $z^A$ (with $A=1\dots, n$) denote complex scalar fields, and 
${F}^{\pm I \, \mu\nu}$
(with $I=0,\dots ,n$) are the (anti-)selfdual abelian field strengths
(including the graviphoton field strength).
An intrinsic definition of a special K\"ahler manifold \cite{c} can be given
in terms of a flat $(2n+2)$-dimensional symplectic
bundle over the $(2n)$-dimensional K\"ahler-Hodge manifold, with 
the covariantly holomorphic sections
\beqa
V=\pmatrix{X^I\cr F_I},\label{section} 
\eeqa
obeying the symplectic constraint
\beqa
i\langle \bar V, V\rangle =i (\bar X^I F_I-\bar F_I X^I)=1\, 
.\label{constr} 
\eeqa
Usually the $F_I$ can be expressed in terms of a holomorphic 
prepotential $F(X)$, homogenous of degree
two, via $F_I=\partial F(X)/\partial X^I$. 
The field-dependent gauge couplings in (\ref{action}) can then 
also be expressed in terms of derivatives of $F$.
The constraint (\ref{constr}) can be solved by introducing 
the projective holomorphic sections $X^I(z)$, which are related to the $X^I$ 
according to 
\beqa
X^I =e^{{1\over 2}K(z,\bar z)}X^I(z), 
\quad K(z,\bar z)=-\log\lbrack i\bar X^I(z)F_I(X^I(z))
-iX^I(z)\bar F_I(\bar X^I(\bar z))\rbrack \,.\label{kp}
\eeqa
Here $K(z,\bar z)$ is the K\"ahler potential which gives rise to the 
metric $g_{A\bar B}$. The holomorphic sections transform under 
projective transformations $X^I(z)\to \exp[f(z)]\,X^I(z)$, which 
induce a K\"ahler transformation on the K\"ahler potential $K$ and a 
U(1) transformation on the section $V$, 
\beqa
K(z,\bar z)\rightarrow K(z,\bar z)-f(z)-\bar f(\bar z)\,, \qquad 
V(z,\bar z) \rightarrow {\rm e}^{{1\over 2}(f(z)-\bar f(\bar 
z))}V(z,\bar z)\,.   \label{kweight}
\eeqa
Besides $V$, the magnetic/electric field strengths 
$(F_{\mu\nu}^I, G_{\mu\nu I})$ also consitute a symplectic 
vector. Here $G_{\m\n I}$ is generally defined by $G^+_{\m\n 
I}(x) = -4i g^{-1/2}\, \d S/\d F^{+\m\n I}$. 
Consequently, also the corresponding magnetic/electric charges 
$Q=(p^I, q_I)$ transform as a symplectic vector. 

In terms of 
these symplectic vectors the stationary solutions have been 
discussed in \cite{BehrndtLuestSabra} in a fixed K\"ahler gauge. 
The generalized  
Maxwell equations can be solved in terms of  
$2n+2$ harmonic functions, which therefore also transform as a 
symplectic vector ($m,n=1,2,3$), 
\beqa
F_{mn}^I=\ft12\epsilon_{mnp}\,\partial_p\tilde H^I(r)\quad ,
\quad G_{mn I}=\ft12\epsilon_{mnp}\,\partial_pH_I(r)\, .\label{gaugeh}
\eeqa
Throughout this paper we 
assume that the metric can be brought in the form \cite{tod}
\beqa
ds^2 = - e^{2U} dt^2 + e^{-2 U} dx^m dx^m \ ,\label{metrican}
\eeqa
where $U$ is a function of the radial coordinate $r=\sqrt{x^m x^m}$. 
The harmonic functions can be parametrized as
\beq
\tilde H^I(r)  = \tilde h^I+{p^I\over r}\,, \qquad 
H_I(r)=h_I+{q_I\over r}\,,  \label{harmonic}
\eeq
and we write the corresponding symplectic vector as $H(r)=(\tilde 
H^I(r), H_I(r)) = h + Q/r$. 

Once one has identified the various symplectic vectors that play a 
role in the solutions, it follows from symplectic covariance 
that these vectors should satisfy a certain proportionality relation. The 
simplest possibility is to assume that $V$ and $H$ are directly 
proportional to each other. Because $H$ is real and 
invariant under U(1) transformations, there is a complex 
proportionality factor, which we denote by $Z$. Hence we define
a U(1)-invariant symplectic vector (here we use the homogeneity 
property of the function $F$), 
\beq
\Pi=\bar Z V=(Y^I,  F_I(Y)) \,,\label{newsec}
\eeq
so that $Y^I=\bar ZX^I$, and assume 
\beq
\Pi (r) -\bar \Pi(r) = i H(r)\,. \label{stable}
\eeq
This equation determines $Z$,
\beq
Z(r)= - H_I(r)\, X^I + \tilde H^I(r)\, F_I(X)\,,\qquad \vert 
Z(r)\vert^2 = i \langle 
\bar \Pi(r) ,\Pi(r)\rangle \,.  \label{Z}
\eeq
The first of these equations indicates that $Z(r)$ is related
to the auxiliary field $T_{\mu \nu}^-$.  This relation may not hold when
$F$ depends on $W^2$.
The equations (\ref{stable}), which we
call the stabilization equations, also govern the $r$-dependence of
the scalar moduli fields: $z^A(r)=Y^A(r)/ Y^0(r)$. So the constants
$(\tilde h^I,h_I)$ just determine the asymptotic values of the scalars at
$r=\infty$. In order to obtain an asymptotically flat metric with 
standard normalization, these
constants must fullfill some constraints. Near the horizon 
($r\approx 0$), (\ref{stable})
takes the form used in \cite{BCDWKLM} and $Z$ becomes proportional 
to the holomorphic BPS mass ${\cal M}(z) = q_I X^I(z) - p^I F_I (X(z)$.

When in addition we make the symplectically invariant ansatz 
$e^{-2U} = Z \bar Z$, it can  
be shown that the solution preserves half the supersymmetries, 
except at the horizon and at spatial infinity, where 
supersymmetry is unbroken. {From} the form of the static solution 
at the horizon ($r\rightarrow 0$) we can easily derive its macroscopic
entropy. Specifically the Bekenstein-Hawking entropy is given by 
\beqa
{\cal S}_{\rm BH}&=&\pi \,(r^2e^{-2U})_{r=0}=\pi\,(r^2Z\bar Z)_{r=0}
\nonumber\\
&=&i\pi\,\Big(\bar Y^I_{\rm hor}\,F_I(Y_{\rm hor})-\bar F_I
(\bar Y_{\rm hor})\,Y^I_{\rm hor}\Big)\,,\label{entropy}
\eeqa
where the symplectic vector $\Pi$ at the horizon,
\beqa
\Pi(r)\stackrel{r\to 0}{\approx} {\Pi_{\rm hor}\over r}\,,\qquad 
Y^I(r)\stackrel{r\to 0}{\approx} {Y^I_{\rm hor}\over r} \ ,\label{qhor}
\eeqa
is determined by the following set of stabilization equations:
\beqa
\Pi_{\rm hor}-\bar\Pi_{\rm hor} = i Q.\label{stabhor}
\eeqa
So we see that the entropy as well as the scalar fields $z^A_{\rm hor}=
Y^A_{\rm hor}/ Y^0_{\rm hor}$ depend only on the magnetic/electric
charges $(p^I,q_I)$. It is useful to note that 
the set of stabilization equations (\ref{stabhor})
is equivalent to the minimization of $Z$ with respect to
the moduli fields \cite{FerrKallStro}. As already said, at the 
horizon $r=0$ full $N=2$ supersymmetry is 
recovered. Also note that in the same way one can construct more general
stationary solutions, such as rotating $N=2$ black holes, multi-centered
black holes, TAUB-NUT spaces, etc. \cite{BehrndtLuestSabra}.

As an example, consider a type-IIA compactification on a Calabi-Yau
3-fold. The number of vector superfields is given as $n =h^{(1,1)}$.
The prepotential, which is purely classical, contains the
Calabi-Yau intersection numbers of the 4-cycles, $C_{ABC}$, 
and, as $\alpha'$-corrections, the Euler
number $\chi$ and the rational instanton numbers $n^r$. Hence the
black-hole solutions will depend in general on all these topological
quantities \cite{BCDWKLM}. 
However, for a large Calabi-Yau volume, 
only the
part from the intersection numbers survives and the $N=2$ prepotential
is given by 
\beqa
F(Y)=D_{ABC}{Y^AY^BY^C\over Y^0},\qquad D_{ABC}=- \ft16 C_{ABC}\ .
\label{cprep}
\eeqa
Based on this prepotential we consider in the following 
a class of non-axionic
black-hole solutions (that is, solutions with purely imaginary moduli fields) 
with only
non-vanishing charges $q_0$ and $p^A$ ($A=1,\dots ,h^{(1,1)}$). So
only the harmonic functions $H_0(r)$ and $\tilde H^A(r)$ are nonvanishing.
This charged configuration corresponds, in the type-IIA 
compactification, 
to the intersection of three D4-branes, wrapped over the internal Calabi-Yau 
4-cycles and hence carrying magnetic charges $p^A$, plus one D0-brane
with electric charge $q_0$. In the corresponding M-theory picture
these black holes originate from the wrapping of three M5-branes, 
intersecting over a common string, plus an M-theory wave solution 
(momentum along the common string).
For the solution indicated above, the four-dimensional metric of the extremal
black-hole solutions is given by \cite{BehrndtLuestSabra}
\beqa
e^{-2U(r)}=2\sqrt{H_0(r)\,D_{ABC}\,\tilde H^A(r)\,\tilde H^B(r)\,\tilde 
H^C(r)} .
\label{IIAmetric}
\eeqa
The scalars at the horizon are determined as
\beqa
z^A_{\rm hor}={Y^A_{\rm hor}\over Y^0_{\rm hor}}\,,\quad
Y^A_{\rm hor}=\ft12 i p^A, \quad Y^0_{\rm hor}=\ft12 
\sqrt{D\over q_0}\,,\quad D=D_{ABC}\,p^Ap^Bp^C\, .\label{IIAscal} 
\eeqa
Finally, the corresponding macroscopic entropy takes the form 
\cite{BCDWKLM,BehCaKaMo}
\beqa
{\cal S}_{\rm BH}=2\pi\sqrt{q_0D}.\label{IIAentropy}
\eeqa

Let us now turn to the discussion of $N=2$ black-hole solutions in the
presence of higher-derivative terms in the $N=2$ supergravity action.
In that situation the function $F$ depends on both $X$ and $W^2$ and is still 
holomorphic and homogeneous of second degree. 
Just like the chiral superfield, whose lowest component is $X^I$, 
$W_{\m\n}$ is also a 
reduced chiral superfield.  Therefore it has the same  
chiral and Weyl weights as $X^I$.  Thus, 
 $W^2= W_{\m\n}W^{\m\n}$ is a scalar chiral  
multiplet with Weyl and chiral weight equal to twice that of the 
$X^I$.  Hence the homogeneity of $F$ implies
\beq 
X^I \,F_I(X,W^2) + 2 \,W^2 \,{\pa F(X,W^2) \over \pa W^2} = 2 F(X,W^2) \,.
\eeq
The lowest-$\theta$ component of the 
superfield $W$ contains the auxiliary tensor field that 
previously took the form of the graviphoton field strength (up to 
fermionic terms). However, in the case at hand the Lagrangian is 
more complicated, so  
that this tensor can only 
be evaluated as a power series in terms of external momenta 
divided by the Planck mass. Similar comments apply to all the 
auxiliary fields. Nevertheless we can still use the 
superconformal $N=2$ multiplet calculus 
\cite{BergshoeffdeRoodeWit}. Obviously, this case is much more 
complicated and we do not attempt to give a full treatment of 
the solutions here.  A detailed discussion of these solutions will appear
elsewhere.  In the following we will instead rely on symplectic covariance
to analyze 
the 
immediate consequences of the $W$-dependence of $F$ 
and discuss its implication for the black-hole entropy.

Our strategy will be to 
expand
$F(X, W^2)$ as a power series in $W^2$ as follows,
\beqa
F(X^I,W^2)=\sum_{g=0}^\infty F^{(g)}(X^I)W^{2g}\,,\label{fg}
\eeqa
where $F^{(0)}$ is nothing else than the prepotential discussed before.
This expansion will allow us to make contact with the microscopic
results of \cite{msw,vafa}, where a suitable expansion of the microscopic 
entropy was proposed.
{From} the fact that the superfield $W$ contains the Weyl tensor at 
order $\theta^2$, the $W^2$ dependence leads, for instance, to 
terms proportional to the square of the Weyl tensor and quadratic 
in the abelian field strengths, times powers 
of the tensor field $T$.  
Nevertheless, these terms are all precisely encoded in 
(\ref{fg}). 

For type-II compactifications on a Calabi-Yau space, the terms involving 
$F^{(g)}(X)$ arise at $g$-loop order, whereas in the dual heterotic vacua
the $F^{(g)}$ appear at one loop and also contain 
non-perturbative corrections. In passing we note that the 
physical couplings are in general non-holomorphic, where the 
holomorphic anomalies are governed by a set of recursive holomorphic anomaly
equations \cite{AntonGavaNarTayl,Bershadskyetal,DWCLMR}.

The presence of the chiral background $W^2$ will not modify the 
special geometry features that we discussed before, provided one 
now uses the {\it full} function $F$ with the $W^2$ dependence 
included. So, the section $V$, which transforms as a
vector under Sp$(2n+2)$, now takes the form
\beqa
V=\pmatrix{X^I\cr F_I(X,W^2)}=\pmatrix{X^I\cr \sum_{g=0}
F^{(g)}_I(X)\, W^{2g}}\,.\label{sectionw} 
\eeqa
In order for the Einstein term to
be canonical, $V$ has to obey again the symplectic constraint (\ref{constr}),
so that 
$\bar X^I F_I(X,W^2)-X^I\bar F_I(\bar X,
\bar W^2)= -i$.
Using the U(1)-invariant combinations $Y^I=\bar ZX^I$ and
$\bar Z^2W^2$ we can define a U(1)-invariant symplectic vector as
\beqa
\Pi=\bar Z V=\pmatrix{Y^I\cr F_I(Y,\bar Z^2W^2)}
=\pmatrix{Y^I\cr \sum_{g=0}F^{(g)}_I(Y) \, (\bar Z^2W^2)^g}\,,\label{newsecw}
\eeqa
where we used the expansion $F(Y,\bar Z^2W^2)=\sum_{g=0}F^{(g)}(Y)(\bar Z^2
W^2)^g$. The factor $Z$ will again be determined by the 
stabilization equations and will thus depend on $W^2$. 

Second,  the  field strength $G_{\mu\nu I}$, which 
together with $F_{\mu\nu}^I$ forms a symplectic pair, is still 
defined by the derivative with respect to the abelian field 
strength of the full action, and therefore modified by the 
$W$-dependence of $F$. Prior to eliminating the 
auxiliary fields, this action is at most quadratic in the 
field strengths, and
$G_{\m\n I}^\pm$ is generally parametrized as 
\beqa
G^+_{\mu\nu I}=\bar F_{IJ}(\bar X,\bar W^2)F^{+J}_{\mu\nu} + {\cal O}_{\mu\nu 
I}^+\,, \qquad G^-_{\mu\nu I}= F_{IJ}(X, W^2)F^{-J}_{\mu\nu} + 
{\cal O}_{\mu\nu I}^- \,, \label{defG}
\eeqa
where 
${\cal
O}^\pm_{\mu\nu I}$ represents bosonic and fermionic 
moment couplings to the vector fields, such that the 
Bianchi identities and the field equations read  
$\partial^\nu(F^+- F^-)_{\mu\nu}^I = \partial^\nu(G^+- G^-)_{\mu\nu I} =0$.
Again, it is crucial to include the full dependence on the Weyl
multiplet, also in the tensors $G^\pm_{\mu\nu I}$ and ${\cal
O}^\pm_{\mu\nu I}$. The reason is that the symplectic
reparametrizations are linked to the full 
equations of motion for the vector 
fields (which involve the Weyl multiplet) and not to (parts of)
the Lagrangian. The modification of the field strength $G_{\mu\nu 
I}$ can be interpreted
as having the 
effect that the electric charges $q_I$ (which, together with
the magnetic charges $p^I$ form the symplectic charge vector $Q$)
get modified in the chiral $W^2$ ``medium",
compared to their original ``microscopic'' values $q^{(0)}_I$.
Below we will comment on the relevance of this observation.

Extremal $N=2$ black-hole solutions in the presence of higher-derivative
interactions must again preserve half the supersymmetries, except 
for $r=0,\infty$, but obviously this condition is now much harder 
to solve. Nevertheless it is possible to show that the (tangent-space) 
derivative of the moduli with respect to the radial variable is still 
vanishing 
at the horizon, 
indicating the expected fixed-point behaviour. 
Rather than solving the equations for the full black-hole 
solution, we impose the stabilization equations. For the metric 
we make again the ansatz (\ref{metrican}) where
$e^{-2U}$ takes the same form in terms of $\Pi$, which now 
incorporates the modifications due to the background,  
\beqa
e^{-2U}&=&Z\bar Z=i(\bar Y^IF_I(Y,\bar Z^2W^2)-Y^I\bar F_I(\bar Y,
Z^2\bar W^2))
.\label{metricw}
\eeqa
The stabilization equations now read 
\beqa
\pmatrix{Y^I-\bar Y^I\cr F_I(Y,\bar Z^2W^2)-\bar F_I(\bar Y,Z^2\bar W^2)}=
i \pmatrix{\tilde H^I(r)\cr H_I(r)}=
i \pmatrix{\tilde h^I\cr h_I} + {i\over r} \pmatrix{p^I\cr q_I}\,
,\label{solutw} 
\eeqa
where the harmonic functions characterize the values of the field 
strengths according to (\ref{gaugeh}), just as 
before, except that the field strength $G_{\m\n I}$ incorporates 
the modifications due to the background.

We are now particularly interested in the behavior of our solution at the
horizon $r\rightarrow 0$, that is we would like to compute the corrected 
expression
for the black-hole entropy. Without any $W$-dependence in $F$,   
previous calculations 
show that 
the quantity ${\bar Z}^2 W^2=\bar Z^2(T_{\mu\nu}^-T^{-\mu\nu})$ 
has the following behaviour near the horizon:
\beqa
{\bar Z}^2 W^2 =  \frac{1}{r^2}  + {\cal O}(r^0) \;\;\;.\label{zerow}
\eeqa
Equation (\ref{zerow}) 
could get modified
when $F$ depends on $W$. These
corrections should then 
be viewed as the back reaction of the non-trivial 
$W^2$-background on
the black-hole solution.
We will work to leading order and 
therefore 
assume that possible corrections to (\ref{zerow})
can be neglected to that order.
Thus, we will simply use equation (\ref{zerow}) at the horizon, so
that the modified $N=2$ black-hole entropy is then given by:
\beqa
{\cal S}_{\rm BH}&=&\pi\,(r^2e^{-2U})_{r=0}=\pi\, (r^2Z\bar Z)_{r=0}
\nonumber\\
& =&
 i \pi \left( 
{\bar Y}_{\rm hor}^I\, F_I(Y_{\rm hor}, 1) - Y^I_{\rm hor} \, {\bar F}_I({\bar 
Y}_{\rm hor}, 1)  \right) \, .
\label{entro}
\eeqa
As before, the symplectic vector $\Pi_{\rm hor}$ is given as in
equation (\ref{stabhor}), where now, however, $\Pi_{\rm hor}$ has a non-trivial
dependence on the $W^2$-background.  So we have 
\beqa
Y^I_{\rm hor}-{\bar Y}^I_{\rm hor}=i p^I,\qquad 
F_I(Y_{\rm hor},1)-\bar F_I({\bar Y}_{\rm hor},1)= iq_I\, .  
\label{stabilw} 
\eeqa
Consequently, the modified black hole 
entropy only depends again on the magnetic/electric charges.

For concreteness, 
let us again discuss the type-IIA compactification on a Calabi-Yau 3-fold
in the limit of large radii, which amounts to suppressing all 
$\alpha'$-corrections.  We are, in particular, 
interested in the contribution from $F^{(1)}$, which
arises at one loop in the type-IIA string. This term is of topological 
origin;
it is related 
to a one-loop $R^4$ term
in the ten-dimensional effective IIA action \cite{Mina}. 
For
the function $F(Y,\bar Z^2W^2)$ we thus take
\beqa
F(Y, {\bar Z}^2 W^2) = F^{(0)}(Y) + F^{(1)}(Y) \bar Z^2 W^2 =
D_{ABC} \, \frac{ Y^A Y^BY^C}{Y^0} 
-\ft1{24} c_{2A}\, \frac{Y^A}{Y^0} \, {\bar Z}^2 W^2 
\;\;\;.\label{ourf0f1}
\eeqa
Here the $c_{2A}$ are the second Chern class 
numbers of the Calabi-Yau 3-fold. For simplicity we will
consider again axion-free black holes
with $p^0 =0, q_A =0$.  
Then the stabilization equations (\ref{stabilw})  have the solution
\beqa
Y^A_{\rm hor} = \ft12 {i} p^A \,, \qquad 
Y^0_{\rm hor} =  \ft{1}{2} \sqrt{\frac{D}{q_0}}\, 
\sqrt{1+\frac{c_{2A}\, p^A }{6D} }\, .
\label{yhorf}
\eeqa
Inserting (\ref{yhorf}) into (\ref{entro}) yields
\beqa
{\cal S}_{\rm BH} 
=  2 \pi \sqrt{q_0} \, \frac{D + {1\over 12}  c_{2A}\,p^A}
{\sqrt{D + {1\over 6}  c_{2A}\,p^A}} \;\;\;.\label{wrong}
\eeqa
Equation (\ref{wrong}) is only to be trusted to linear order
in $c_{2A}$, because we have not included the back reaction.
Expanding (\ref{wrong}) to lowest order in $c_{2A}$ yields
\beqa
{\cal S}_{\rm BH} = 2 \pi \sqrt{q_0 D} + 2 \pi ( \ft{1}{12} - \ft{1}{12})
c_{2A}\, p^A\,  \sqrt{ \frac{q_0}{D} }  + \cdots\,.
\label{mentro}
\eeqa
The terms linear in $c_{2A}$ thus cancel out!  
Thus,
when expressed in terms of the charges $Q=(p^I,q_I)$, 
there is no correction to ${\cal S}_{\rm BH}$ to lowest order in $c_{2A}$!.
Although this is a rather striking result, whose significance
is not quite clear to us at the moment, it seems to disagree 
with the microscopic findings of \cite{msw,vafa}, as we will now discuss.

We would like to compare the macroscopic entropy formula (\ref{mentro})
with the microscopic entropy formula recently computed in \cite{msw,vafa}
for certain 
compactifications of M-theory and type-IIA theory on Calabi-Yau
3-folds.  In \cite{msw,vafa} the microscopic entropy was,
to all orders in $c_{2A}$, found to be
\beqa
{\cal S}_{\rm micro}=2\pi\sqrt{q_0
D\Big(1+{c_{2A}\,p^A\over 6D}\Big)}\, .\label{malmicro}
\eeqa
Expanding (\ref{malmicro}) to lowest order in $c_{2A}$ yields
\beqa
{\cal S}_{\rm micro}=2\pi\sqrt{q_0 D}+
2 \pi 
\ft1{12}c_{2A}\,p^A
\sqrt{{q_0\over D}}  + \cdots \ .\label{entr1msw}
\eeqa
The correction in (\ref{entr1msw}) was then matched \cite{msw,vafa}
with a correction to the effective action involving $R^2$-type terms
with a coefficient function $F^{(1)}$.

The approach used above to obtain the macroscopic entropy (\ref{mentro})
is different from the approach used in \cite{msw,vafa} for obtaining
the macroscopic formula.  
A  comparison of the results of both approaches, that is of
(\ref{entr1msw}) and (\ref{mentro}),  indicates that
in order to obtain matching of both results, the 
macroscopic
electric charge $q_0$ appearing in (\ref{mentro}) cannot be identical
to the electric $q_0$ appearing in (\ref{entr1msw}).  This is one way
of explaining the discrepancy.
If we denote the
charge $q_0$ appearing in (\ref{entr1msw}) by $q_0^{(0)}$, then 
matching of (\ref{entr1msw}) and (\ref{mentro}) can be achieved provided
that the $q_0$ appearing in (\ref{mentro})
is related to $q_0^{(0)}$ as follows:
\beqa
q_0 = q_0^{(0)} 
\left(1+{c_{2A}p^A\over 6D}\right) \;\;\;.
\label{mimaq}
\eeqa  
As alluded to earlier, this can be interpreted as a modification
of the ``microscopic'' charge  $q_0^{(0)}$ due to the $W^2$ medium.
This interpretation can be further motivated by noting that
if one defines $q_0^{(0)}$ to be given by
\beqa
q_I^{(0)} \equiv -i \left( F_I^{(0)}(Y_{\rm hor}) - 
{\bar F}_I^{(0)}({\bar Y}_{\rm hor}) \right)= q^I + i
\sum_{g \geq 1} \left( F_I^{(g)}(Y_{\rm hor}) - 
{\bar F}_I^{(g)}({\bar Y}_{\rm hor}) \right) \;\;\;,
\label{qmicro}
\eeqa
then insertion of (\ref{yhorf}) into (\ref{qmicro}) precisely
yields (\ref{mimaq}).  Equation (\ref{qmicro}) makes it 
clear that, in the presence of a $W^2$ medium, the charges
$q_0$ and $q_0^{(0)}$ cannot be identical.  Since the 
``microscopic'' charge  $q_0^{(0)}$ has the interpretation
of quantized momentum around a circle \cite{msw}, it is
integer valued.  We thus note that the macroscopic charge $q_0$
given in (\ref{mimaq}), which is measured at spatial infinity,
is not integer valued in the presence of a
$W^2$ medium.

Given that the equations (\ref{entr1msw}) and (\ref{mentro}) agree (provided
the relation (\ref{mimaq}) holds), it is then conceivable
that the full macroscopic entropy, derived  
from the metric given in (\ref{metricw}), also
agrees with the full microscopic entropy (\ref{malmicro}).
In order to calculate the full macroscopic entropy from (\ref{metricw}),
one will have to take into account a possible back reaction of the non-trivial
$W^2$ background on the black hole solution, 
in particular the corrections to the quantity
$(\bar Z^2W^2)$ at the horizon (\ref{zerow}). 
In the presence of such
corrections, the quantity $(\bar Z^2W^2)_{\rm hor}$
will presumably also depend on the magnetic/electric charges, that is
$(\bar Z^2W^2)_{\rm hor}={f(p,q)\over r^2}$.
In addition the higher functions $F^{(g)}$ ($g>1$) might also contribute
to the entropy. Finally, even non-holomorphic corrections to the 
higher-derivative effective action might play a role.
At the end, let us note that if one inserts (\ref{mimaq}) into
(\ref{malmicro}),
the microscopic entropy takes again the
very simple form:
\beqa
{\cal S}_{\rm micro}=2\pi\sqrt{q_0D}.
\eeqa
So with this charge ``renormalization" one rediscovers
the zero-th order entropy (\ref{IIAentropy}).
It is tempting to speculate that this 
feature is related to the enhancement
of the supersymmetry at the horizon, namely to the fact that at the horizon
we still have an $AdS_2\times S^2$ geometry, even after including
all higher-derivative terms.

{\large \bf Acknowledgements}

\smallskip
\noindent
This work is supported by 
the Deutsche Forschungsgemeinschaft (DFG) and by the European 
Commission TMR 
programme ERBFMRX-CT96-0045 in which Humboldt-University Berlin
and Utrecht University participate. 
W.A.S is partially supported by DESY-Zeuthen.

{\large \bf Note Added}

In \cite{wald} it has been pointed out that
there are modifications to the
Bekenstein-Hawking
entropy formula
 in the presence
of higher curvature terms.
In this paper we have not considered such modifications.  It is not
clear at present what their contribution to our result is.
We would like to thank Serge Massar
for raising this issue and for bringing references \cite{wald} to
our attention.  B.~d.~W. thanks Robert Myers for useful discussions
regarding this topic.

\end{document}